\RequirePackage{fix-cm}
\documentclass[smallextended]{svjour3}       
\smartqed  
\usepackage{appendix}
\usepackage{graphicx}
\usepackage{amssymb}
\usepackage{color}
\usepackage{amsmath}
\usepackage{cite}
\usepackage{multirow}
\usepackage{float}
\newcommand{\etal}{  {\it et al.}}

\newcommand{\jastp}{  {\it J. Atmos. Solar-Terr. Phys.}}

\begin{document}

\title{Computer-aided measurement \\of the heliographic coordinates of sunspot groups 
}

\author{H. \c{C}akmak}

\titlerunning{Computer-aided measurement of the heliographic coordinates of sunspot groups}

\authorrunning{H. \c{C}akmak} 

\institute{H. \c{C}akmak\at
  Istanbul University Science Faculty, Astronomy and Space Science Department \\
  34119 Beyaz{\i}t / Istanbul - Turkey\\
  \email{hcakmak@istanbul.edu.tr}           
}

\date{Received: date / Accepted: date}

\maketitle

\begin{abstract}

Heliographic coordinates are used to identify the positions of solar features, especially the sunspot groups on the Sun's surface. Tracking the positions of sunspot groups provides information about solar rotation and the movement behavior of sunspot groups over Solar Cycles. Two heliographic coordinates are defined: Carrington and Stonyhurst. The calculations used here depend on three solar parameters, position angle {\it P}, latitude angle $\it {B_0}$ and the starting longitude $\it {L_0}$. These values are calculated by using {\it Astronomical Almanac} for the observation time. In this study, a computer program called {\it Computer Aided Measurement for Sunspots} (CAMS) is presented. The main aim of the program is to determine the heliographic latitude and longitude of the sunspot groups besides other features like the latitudinal and longitudinal length, rectangular area and the tilt angle. This is accomplished by generating the corresponding Stonyhurst disk for the {\it P} and $\it {B_0}$ angles of the time of observation and superimposing it onto the scanned drawing. Since 2009, CAMS is being used to process daily solar drawings at the Istanbul University Observatory.

\end{abstract}

\section{Introduction}
     \label{S-Introduction} 
Sunspots are the first noticeable feature on the visible surface of the Sun. The number of spots on the Sun's surface increase and decrease during a specific time range which is 11 years on average. This period is known as Solar Cycle or Sunspot Cycle. During the minimum phase of a cycle, no sunspots are seen, but, at the maximum phase, many sunspots can be observed on the Sun's surface (Fig. 1). Sometimes big-sized spots are visible to the naked-eye under proper conditions. These sunspots have attracted attention through the centuries. Therefore, since the 17th century, sunspot observations are being continued without interruption. 

\begin{figure}[t]    
   \centerline{
	\includegraphics[width=1.0\textwidth,clip=]{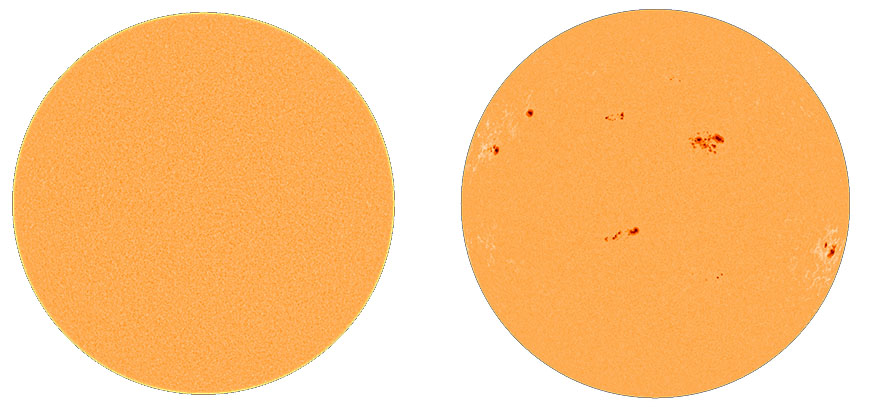}
   }
   \caption{Two typical views of a Sunspot Cycle. $\it Left$: Minimum, $\it Right$: maximum phase. Images were 
taken from the SOHO (Solar and Heliospheric Observatory) archive.}
   \label{F-figures-01}
\end{figure}
One of the most known Sun observers, Galileo showed that movement of the sunspots across the apparent disk of the Sun is similar to rotating across a spherical surface, rather than a flat disk. This concept made it possible to develop a spherical coordinate system of solar latitude and longitude, similar to the latitude and longitude system on Earth. Today, two heliographic coordinates are defined: Carrington and Stonyhurst. They differ on whether the lines of longitude are fixed for the Sun's rotation (in Carrington) or fixed for the observer (in Stonyhurst) \cite{Carrington1863, Teague1996, Thompson2006, SanchezBajo2013}. More explanations about both heliographic coordinates can be found in the articles of Thompson\cite{Thompson2006} and {Sanchez-Bajo \& Vaquero} \cite{SanchezBajo2013}.  

More properties about the positions of the sunspot groups were discovered with increasing solar observations in the past. German astronomer Christoph Scheiner showed that the solar rotation rate varied with latitude. This meant that the Sun does not rotate as a solid body. Carrington published his observations of sunspot positions in 1863. And, he estimated the correlation between solar rotation rate and latitude\cite{SanchezBajo2013}. In 1904, Edward Maunder showed the latitude of sunspot emergence changes from the beginning to the end of the sunspot cycle in a diagram called {\it butterfly}. All these movement behaviors of sunspot groups provide information about the Sun: e.g. the rotation period of the Sun, the nature of the solar cycle and the existence of differential rotation.

Most of observatories spread around the world are continuing to draw the sunspots manually. But, on the other hand, technological developments are making it also necessary to process these sunspot drawings in computer environment. Some of the observatories transferred their old drawings into digital environment via scanners to process them later. Some studies are done on determining the positions of sunspot groups from sunspot drawings. The most known applications are the Helio programs developed by Peter Meadows\cite{Meadows2002}. These are sufficient in some extents, but not adequate for our own needs. Moreover, similar programs are developed by different people: the program HSUNSPOTS is developed by Cristo\etal \cite{Cristo2011} to analyze ancient solar drawings and the program DigiSun is used in the SIDC of the Royal Observatory of Belgium\cite{Clette2011}. 

The CAMS program which will be introduced here has been developed with a different perspective from the others. It aims to obtain as many features of the sunspot groups as possible. Moreover, not only the scanned solar drawings but also CCD and solar digital images can be used with the program by adjusting their size and resolution properly. In this program, a Stonyhurst disk overlaying technique is used. This technique gives the opportunity for calculating most of the properties of the sunspot groups easily. All operations are carried out in two simple steps, superimposing Stonyhurst disk onto image and, marking and classifying the sunspot groups separately. Most of the other processes are performed automatically such as calculating the solar parameters at the time of observation, the heliographic coordinates and other measurable parameters of the sunspot groups and, altering and sorting the groups by their latitude and longitude.

In the sunspot observations, the appearance of the Stonyhurst disk depends mainly on the {\it P} and $\it {B_0}$ angles. Therefore, the calculations of these solar parameters have a special importance and are introduced in detail in Section 2. The algorithm for drawing the Stonyhurst disks is explained in Section 3. And, the basics of CAMS's workings are given in Section 4. The advantages, importance and possible future developments of the CAMS are presented in the Discussion Section.  

\section{Calculations of the Solar Parameters} 
  \label{S-Sun-param}
Any coordinate system used in astronomy must be set up using a plane and a starting point. For heliographic coordinates the plane in question is the equator of the Sun, but the starting point is a little difficult to describe, because there are no permanent features on the surface of the Sun,  which can be taken as a fixed point. Carrington defined this virtual point as a point occupied by the ascending node of the solar equator on the ecliptic at noon on January 1st 1854. By this assumption, heliographic coordinates of a point on the surface of the Sun are defined by the spherical definitions as shown in Fig. 2a\cite{Duffett1988}.
\begin{figure}[h]    
  \label{F-figures-02}
  \centerline{
	\includegraphics[width=1.0\textwidth,clip=]{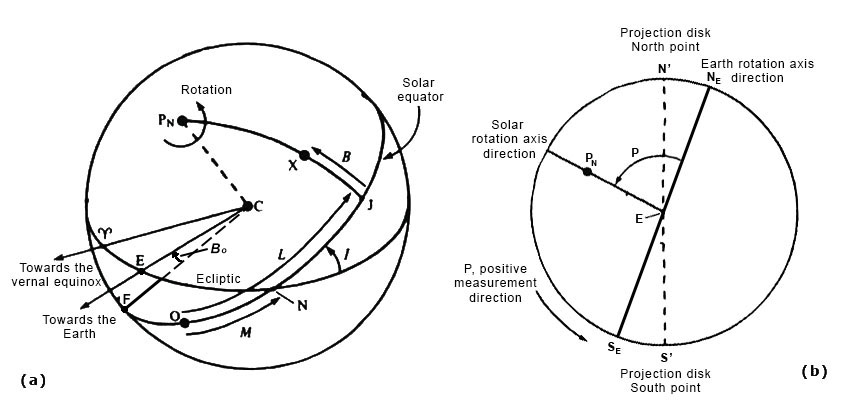}
  }
  \caption{(a) Definition of the heliographic coordinates of a point {\it X} on the surface of the Sun according to the spherical definitions. (b) Position of the rotation axis of the Sun with respect to the rotation axis of the Earth and north $\&$ south line of the solar projection disk.}
\end{figure}

The line of sight is shown by the line towards the Earth in Fig. 2a and it shows the direction to the center of the apparent solar disk, which is labeled as {\it E} in Fig. 2b. Also, the rotation axis of the Sun is shown by {\it E$P_N$} line and the angle between this line and the rotation axis of the Earth is {\it P}, the position angle of the Sun. The angle $\widehat{FCE}$ is the latitude angle {\it $B_o$} and represents the angular distance of the Sun's equator from the center of the apparent solar disk. In accordance with the definitions as shown in Fig. 2a, the starting longitude {\it $L_o$} (point {\it O}) and {\it $B_o$} are calculated by \cite{SmartGreen1977}
\begin{eqnarray*}  \label{Eq-solar-param}
  \begin{aligned}
	L_o &= \arctan \left( \tan (\lambda_{\odot} - \Omega) \cos I \right) + M ,\\[4pt]
	B_o &= \arcsin \left( \sin (\lambda_{\odot} - \Omega) \sin I \right) ,
  \end{aligned} 
\end{eqnarray*}
where $\Omega$ is the angle between the ascending node of the Sun's equator on the ecliptic and the point of the vernal equinox ($\Upsilon$N angle), $\lambda_{\odot}$ is the ecliptical longitude of the Sun, {\it I} is the inclination of the Sun's equator with respect to the ecliptic plane and its value is 7.25$^\circ$. {\it M} is the angular distance of the ascending node of the Sun's equator on the ecliptic from the starting longitude. Here, {\it M} is given by \cite{Duffett1988}
\begin{eqnarray*} \label{Eq-solar-m}
  \begin{aligned}
      M = 360 - M^\prime
  \end{aligned} 
  \quad \quad \text {with} \quad \quad
  \begin{aligned}
      M^\prime &= \frac{360}{25.38} (JD - 2 398 220.0)
 \end{aligned}
\end{eqnarray*}
where JD is the Julian date for the day in question, 2 398 220 is the Julian date at noon on 1 January 1854. {\it M} $^\prime$ must be reduced to the range 0 - 360 degrees.

{\it P} is the angle between both rotation axis directions of the Sun and the Earth. But, this cannot be calculated directly. Therefore, both axis inclinations with respect to ecliptic are calculated separately and {\it P} is the sum of both. So, let $\theta_1$ be the inclination of the Earth's axis and let $\theta_2$ be the inclination of the Sun's axis. And, {\it P} is given by \cite{SmartGreen1977}
\begin{eqnarray*} \label{Eq-solar-teta}
  \begin{aligned}
	\quad  P = \theta_1 + \theta_2
  \end{aligned} 
  \quad \qquad \text{with} \quad \qquad
  \begin{aligned}
	\theta_1 &= \arctan ( -\cos  \lambda_{\odot} \tan \varepsilon ) ,\\[4pt]
	\theta_2 &= \arctan \{ -\cos (\lambda_{\odot} - \Omega)\tan I \} ,
  \end{aligned}
\end{eqnarray*}
where $\varepsilon$ is the angular distance of the Earth's rotation axis from the ecliptic plane. More detailed explanations are given in the {\it Textbook on Spherical Astronomy} of {Smart \& Green} \cite{SmartGreen1977}. 

In general, $\Omega$, $\lambda_{\odot}$ and $\varepsilon$ are changing with time. Therefore, these parameters should also be calculated at the time under consideration and they are given by \cite{Duffett1988}
\begin{eqnarray*} \label{Eq-time-param}
  \begin{aligned}
	\Omega &= 75^\circ 46^\prime + 84^\prime T ,\\[4pt]
	\varepsilon &= 23^\circ 26^\prime 21.45'' - 46.815'' T - 0.0006'' T^2 + 0.00181'' T^3 ,\\[4pt]
	\lambda_{\odot} &= \frac{360}{365.242191}T + \frac{360}{\pi}e \sin \left\{ \frac{360}{365.242191}T 
 + \varepsilon_g - \varpi_g \right\} + \varepsilon_g ,
  \end{aligned} 
\end{eqnarray*}
with
\begin{eqnarray*}
  \begin{aligned}
	\varepsilon_g &= 279.6966778 + 36000.76892\hspace{0.05cm}T + 0.0003025\hspace{0.05cm}T^2 ,\\[4pt]
	\varpi_g &= 281.2208444 + 1.719175\hspace{0.05cm}T + 0.000452778\hspace{0.05cm}T^2 ,\\[4pt]
	e &= 0.01675104 - 0.0000418\hspace{0.05cm}T -0.000000126\hspace{0.05cm}T^2 ,
  \end{aligned}  
\end{eqnarray*}
where {\it T} is the number of Julian centuries since the epoch 2000 January 1.5, $\varepsilon_g$ is the mean ecliptic longitude of the Sun at the epoch, $\varpi_g$ is the longitude of the Sun at perigee and {\it e} is the eccentricity of the Earth's orbit around the Sun.

\section{Preparation of the Stonyhurst disk} 
      \label{S-Stonyhurst}  
In order to form a Stonyhurst disk on an image, it is necessary to form a two-dimensional projection of a three-dimensional sphere. But only half of the sphere or the visible side of the sphere must be taken into consideration. Let this sphere be indicated in Figure 3a. If {\it N} is any point on this sphere and its spherical coordinates in three-dimensional space are the radius of the sphere ({\it r}), latitude angle ($\theta$) and longitude angle ($\varphi$), its Cartesian coordinates ({\it x, y, z}) are given by \cite{GovilPai2004}
\begin{eqnarray*} \label{Eq-cartesian}
  \begin{aligned}
	x &= r \cos \theta \sin \varphi ,\\
	y &= r \sin \theta ,\\
	z &= r \cos \theta \cos \varphi ,
  \end{aligned} 
\end{eqnarray*}
where the angle $\theta$ is selected between -90$^\circ$  and +90$^\circ$ and, the angle $\varphi$ is selected between 0$^\circ$ and 180$^\circ$ \cite{SmartGreen1977}. These three-dimensional (3D) coordinates have to be converted to two-dimensional (2D) coordinates to draw the Stonyhurst disk depending on {\it P} and {\it $B_o$} parameters of the Sun. These are done with the well known transformation and projection equations \cite{WrightSweet2000, GovilPai2004, Stephens2000, Salomon2006}. Let us assume that an angle is rotated in the anti-clockwise direction of its positive and in clockwise direction of its negative values. And also, let us assume that {\it P} and {\it $B_o$} angles are positive. First, the {\it x} axis is rotated {\it $B_o$} degrees and $x_1$, $y_1$, $z_1$ values are obtained, and then, the {\it z} axis is rotated {\it P} degrees and $x_2$, $y_2$, $z_2$ values are obtained (Fig. 3b). These equations are specified as follows \cite{Salomon2006}
\begin{eqnarray*} \label{Eq-rotation}
  \begin{aligned}
	x_1 &= x ,                       &\hspace{1cm} x_2 &= y_1 \sin P + x_1 \cos P ,\\
	y_1 &= z \sin B_o + y \cos B_o , & y_2 &= y_1 \cos P - x_1 \sin P ,\\
	z_1 &= z \cos B_o - y \sin B_o , & z_2 &= z_1 ,
  \end{aligned}
\end{eqnarray*}
Here, $x_2$ and $y_2$ are the 2D projection coordinates of the point {\it N} with respect to the sphere center. When these are rearranged, we get 
\begin{figure}[t]    
  \centerline{
	\includegraphics[width=0.45\textwidth,clip=]{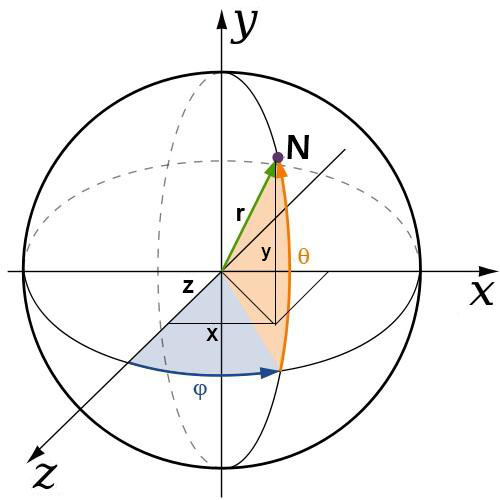}\hspace{0.8cm}
	\includegraphics[width=0.45\textwidth,clip=]{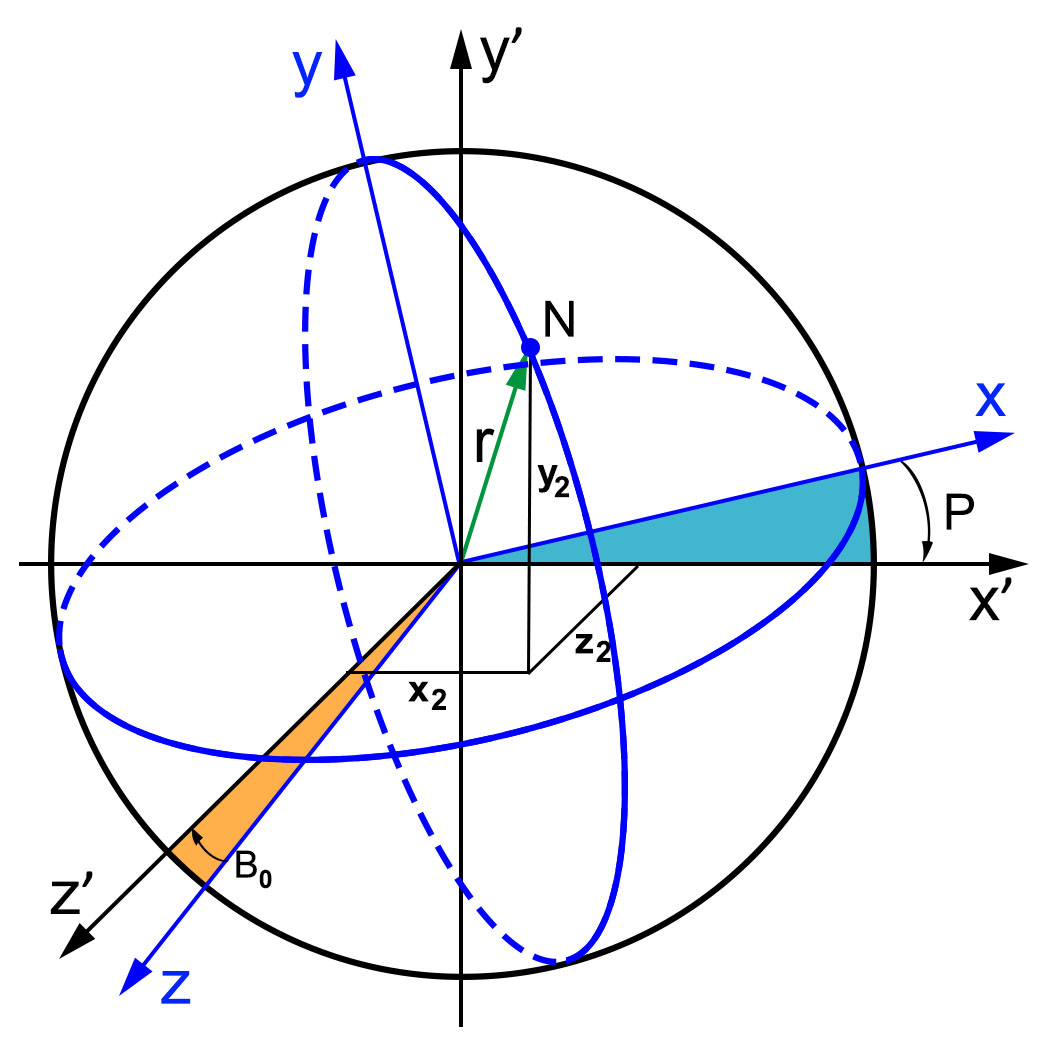}
  }
   \vspace{-0.38\textwidth}
   \hspace{0.045\textwidth}  \color{black} (a)
   \hspace{0.475\textwidth}  \color{black} (b)
   \vspace{0.35\textwidth}
  \caption{\textbf{(a)} Spherical ($\it r, \theta, \varphi$) and Cartesian ($\it {x, y, z}$) coordinates of the point {\it N} in three-dimensional space. \textbf{(b)} Projection coordinates of the point {\it N} are expressed as $\it {x_2, y_2}$ and $\it z_2$ in the $\it {x'y'z'}$ projection system. Here, {\it z} axis (blue) is rotated {\it P} degree and {\it x} axis (blue) is rotated $\it B_0$ degree. Also, blue colored circles represent the rotated planes of the point {\it N}.}
  \label{F-figures-03}
\end{figure}
\begin{eqnarray*} \label{Eq-projection-sp}
	x_2 &=& (z \sin B_o + y \cos B_o) \sin P + x \cos P ,\\
	y_2 &=& (z \sin B_o + y \cos B_o) \cos P - x \sin P ,
\end{eqnarray*}
Since the zero point of the screen coordinates in the computer environments is the top-left corner, the x-axis's value is increasing from the left to right of the screen and, the y-axis's value is increasing from the top to bottom of the screen, Therefore, the center of the any solar disk image on the screen does not coincide with the zero point of the screen coordinates. Hence, in order to draw the sphere onto solar image, the center of the sphere must be moved to the center of the solar disk image. This is done by adding x and y-axis values of the screen coordinates of the center of the solar disk image to the x and y-axis values of the all points of the sphere separately. Let $x_p$ and $y_p$ be the screen coordinates of the point {\it N}. The following equations are obtained with the explanations above:
\begin{eqnarray*} \label{Eq-projection}
  \begin{aligned}
	x_p &= x_0 + x_2,  &\hspace{1cm} x_p &=& x_o + (z \sin B_o + y \cos B_o) \sin P + x \cos P ,\\
	y_p &= y_0 - y_2 , & y_p &=& y_o - (z \sin B_o + y \cos B_o) \cos P + x \sin P ,
  \end{aligned}
\end{eqnarray*}
where {\it $x_0$} and {\it $y_0$} are the coordinates of the disk image center in screen coordinates. The subtraction is carried out in y-values because of the increase in the reverse direction contrary to the Cartesian system. With these projection equations, Stonyhurst disk can be prepared with any values of {\it P} and {\it $B_o$}. As an example, the Stonyhurst disk for {\it $B_o$} = 5$^\circ$ and {\it P} = 0$^\circ$ values is shown in Figure 4a. This is compared with the one used in the observatory (Fig. 4b) and their can be seen agreement in Fig. 4c. Also, other Stonyhurst disks (for {\it $B_o$} = 0,1,2,3,4,5,6,7$^\circ$) were checked as well and exact fits were obtained.
\begin{figure}[h]    
  \centerline{
	\includegraphics[width=1.0\textwidth, clip=]{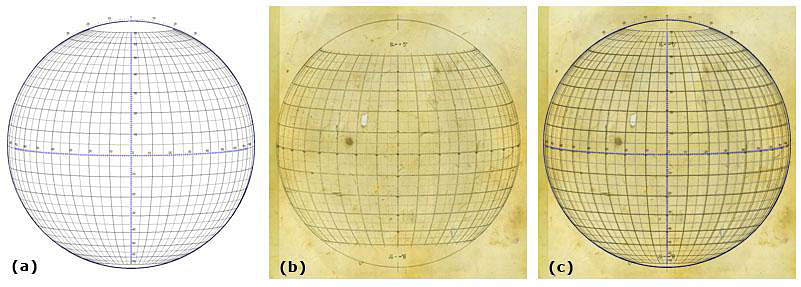}
  }
  \caption{(a) Stonyhurst disk prepared for {\it $B_o$} = 5$^\circ$ in computer. (b) Plastic Stonyhurst disk with the same {\it $B_o$}. (c) Two disks superposed.}
  \label{F-figures-04}
\end{figure}

\section{Computer-aided process of the solar drawings} 
   \label{S-process}
CAMS is mainly designed to process the daily solar drawings, and the interface is presented in Fig. 5. All processes are performed in this window, such as a new process, checking the previous processes, calculation of the solar parameters and transferring the results of the process to the web. Processing the solar drawing has two stages. First stage is the transferring the drawing into the program and superposing the Stonyhurst disk onto it.The second stage is the determining the heliographic coordinates of each sunspot group besides other features like the group length and tilt angle.
\begin{figure}[t]    
  \centerline{
	\includegraphics[width=1.0\textwidth,clip=]{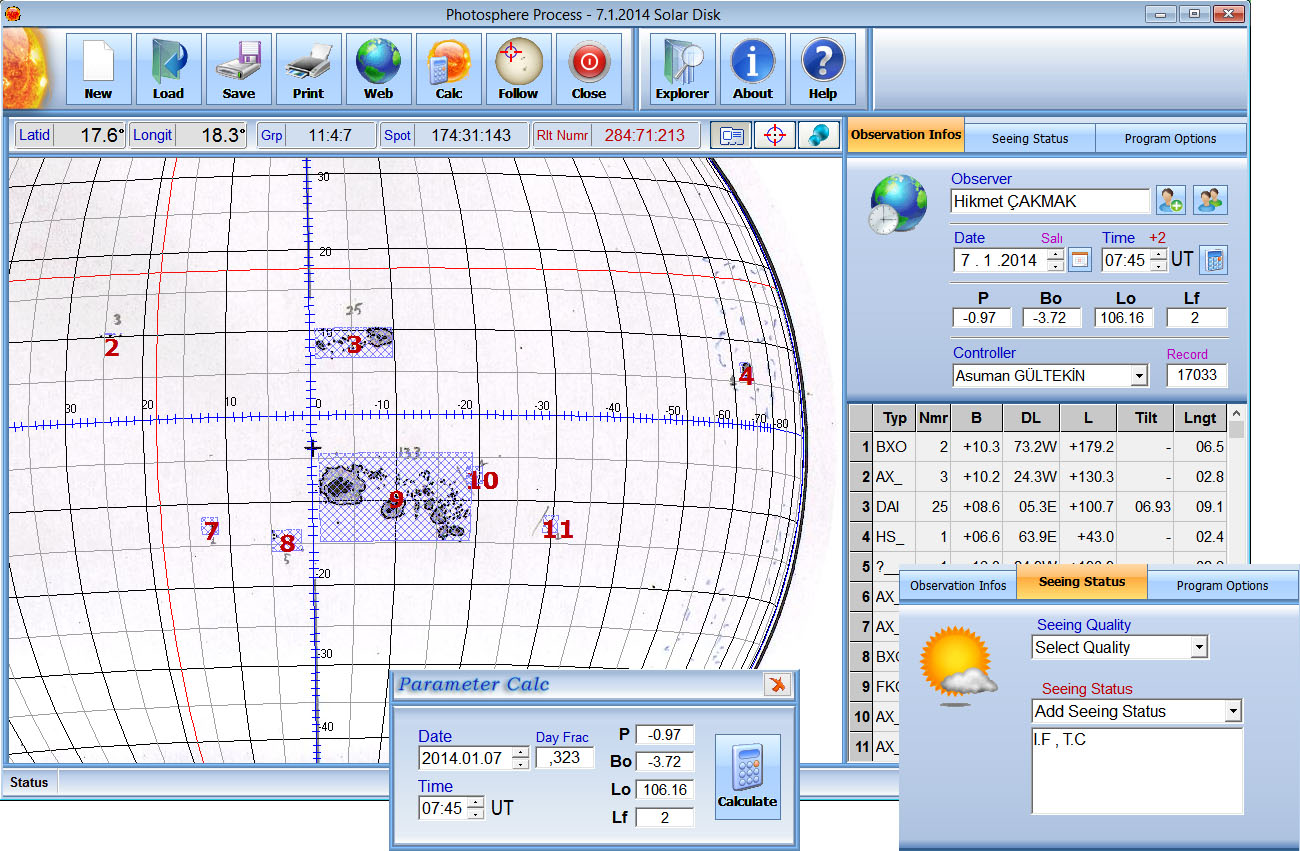}
  }
  \caption{CAMS main window with an example of the transferred sunspot drawing. The solar parameter calculation and image quality selection windows are seen at the bottom, respectively.}
  \label{F-figures-05}
\end{figure}
\begin{figure}[h!]    
  \centerline{ 
	\includegraphics[width=0.8\textwidth,clip=]{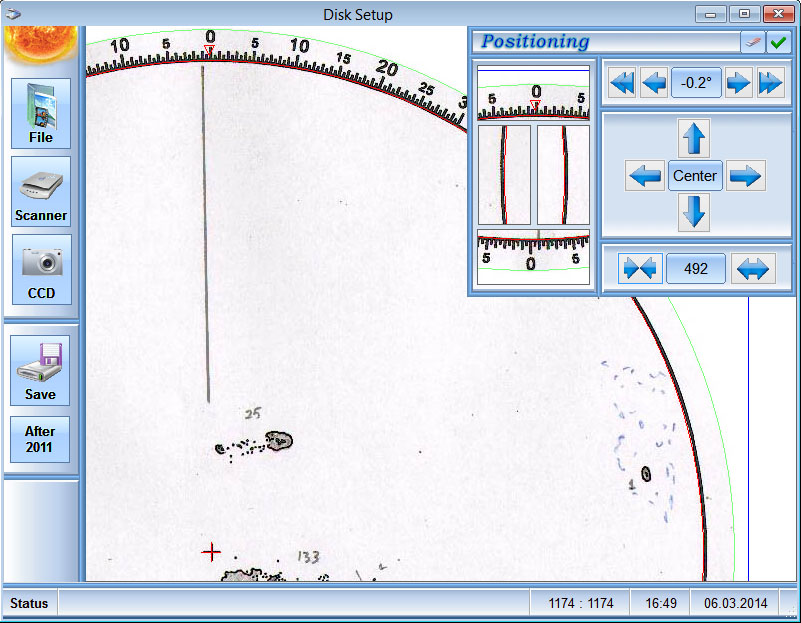}
  }
  \caption{Defining the center and radius of the circle represent the size of the Stonyhurst disk over the image of the drawing.}
  \label{F-figures-06}
\end{figure}

\subsection{Transferring the solar drawing into program and superposing Stonyhurst disk on it} 
   \label{S-process}
After all sunspots on the solar surface are drawn manually on the observation paper, the processing stage starts with the calculation of the solar parameters by entering date and time of the observation in the parameter calculation window. Then, the drawing is transferred into the program by using the scanner interface. After this is done, the center and a point on the outer border of the drawing disk must be marked, respectively. A circle representing the size of the Stonyhurst disk is imposed spontaneously over the image (Fig. 6). Conformity between circle and disk is checked visually by changing the radius and by shifting the center of the circle. Also, the location of the North point of the drawing is adjusted properly by the arrows of the positioning window. After approval of conformity between circle and disk, the cleaning of the outer region and the clipping process of the image is carried out and, the image is transfered into the measurement window.
\begin{figure}[h]    
  \centerline{ 
	\includegraphics[width=0.95\textwidth,clip=]{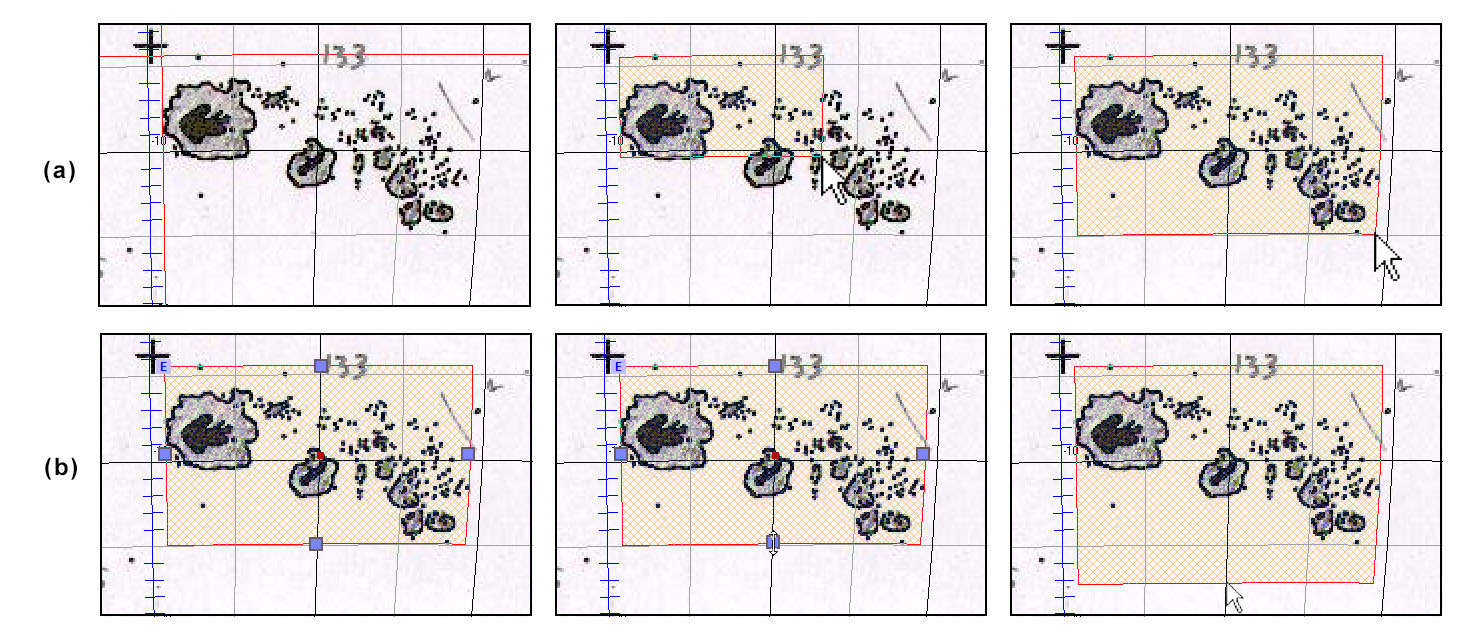}
  }
  \caption{(a) Marking the sunspot group using an enclosed rectangular frame. The beginning, middle and last sequence of the stage. (b) The adjustment can be made easily by moving the handles on the border of the frame.}
  \label{F-figures-07}
\end{figure}

\subsection{Measuring the heliographic coordinate of the sunspot groups} 
   \label{S-measuring}
Stonyhurst disk is drawn automatically over the clipped image of the drawing for the time of observation entered before as seen in Fig. 5. This disk can be redrawn again, if needed, by readjusting the date or the time of the observation on the date \& time section of the program. The measurement process is started by passing big picture mode of the measurement window and the big-sized image of the drawing is loaded into the window. When the mouse moves over the Stonyhurst disk, a spherical cursor will appear and display its position in apparent heliographic coordinates. The sunspot group to be processed is marked as an enclosed rectangular area with this spherical cursor. Some stages of this process are shown in Fig. 7a.

 The rectangular frame's edges should be as narrow as possible while containing the whole group. This proximity can be adjusted more precisely by moving the small handles placed on four edges of the frame, after the marking process is completed (Fig. 7b). Then, a group information window is displayed on top-middle of the measurement window showing all information about the sunspot group. The coordinates at the center of this rectangular frame are the heliographic coordinates of the sunspot group. Its latitudinal and longitudinal width is the height and length of the group, respectively. The type of the group and the number of the sunspots are entered by selecting proper values from the dropdown lists (Fig. 8a). If an A-type group which has a length smaller than 1$^\circ$ is encountered, point source measurement is fulfilled by touching to the center of the group. No length, tilt angle and rectangular area for the group is calculated in this case (Fig. 8b).
\begin{figure}[t]    
  \centerline{ 
	\includegraphics[width=0.8\textwidth,clip=]{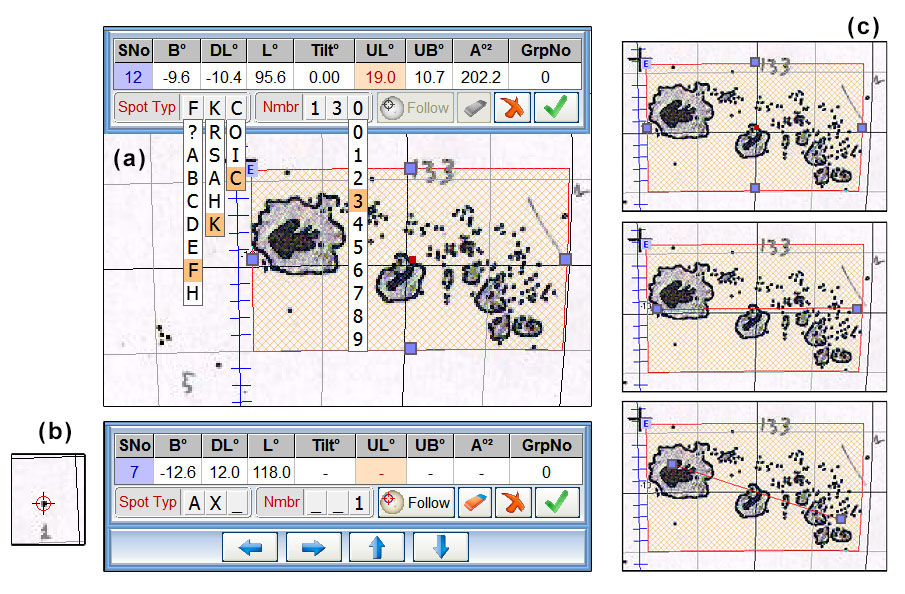}
  }
  \caption{(a) Selecting the type of the group and the number of the sunspot with the drop-down lists. (b) Measuring the point like sunspot and information about it. Fine adjustments are done using the arrow buttons. (c) Positioning the tilt line according to the tilt direction of the group.}
  \label{F-figures-08}
\end{figure}
\begin{figure}[t]    
  \centerline{
	\includegraphics[width=0.8\textwidth,clip=]{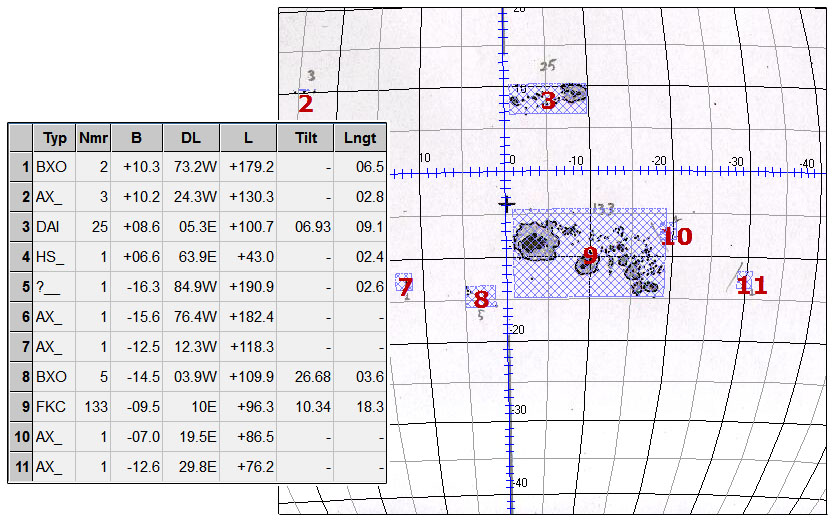}
  }
  \caption{A finished observation process example that all the sunspot groups are measured. {\it Left}: Calculated values for the sunspot groups. {\it Right}: Rectangular frames of the groups with the corresponding sequence number.}
  \label{F-figures-09}
\end{figure}

The tilt angle of the group is started to adjust by pressing the {\it E} symbol on the top-left corner of the group's frame. A line with two small handles at its ends appears in the middle of the frame. Then, the inclination of this line is adjusted by moving these handles in proper position to indicate the tilt direction of the group (Fig. 8c). Adjusting the tilt is concluded by pressing again the {\it E} symbol again. After all the sunspot group are marked and processed individually (Fig. 9), all data about the observation is saved into the database of the program and sent to web page server of the observatory. Detailed information about the CAMS and steps of the processing the observations with CAMS are given in the \c{C}akmak's symposium proceeding\cite{Cakmak2010}.

As an example of the precision of the CAMS, the comparative measurement values of a sample observation day is listed in the Table 1. Image of the solar drawing of this day is shown in Fig. 10. Old-dated solar observation is specially selected to show the capability of the built-in solar ephemeris of the program. In this table, the measurements of the both plastic Stonyhurst disk and CAMS are given for every sunspot group separately. The last two columns of the right panel are the values calculated with CAMS, latitudinal length and tilt angle of the sunspot group, respectively.
\begin{table}[t]
  \renewcommand{\arraystretch}{1.5}
  \setlength{\tabcolsep}{4.7pt}
  \begin{center}
	\caption{The comparative measurement values of a sample observation day. Left panel results are  calculated manually by using the plastic Stonyhurst disk (for $\it B_0$ = 2), whereas right panel results are obtained with CAMS.}
	\begin{tabular*}{1.0\linewidth}{@{}lrrr@{ }lrr|rr@{ }lrrrr}
	  \hline  \hline
	  Date & \multicolumn{2}{l}{19.12.1989} & \multicolumn{3}{r}{P}         & \multicolumn{1}{r|}{+8.06}  & +8.08 \\
	  Time & \multicolumn{2}{l}{09:45 UT}   & \multicolumn{3}{r}{$\it B_0$} & \multicolumn{1}{r|}{-1.51}  & -1.52 \\
	                                          \multicolumn{6}{r}{$\it L_0$} & \multicolumn{1}{r|}{273.50} & 273.47 \\	  
	  \hline  \hline
	Type & S & \multicolumn{1}{c}{B} & \multicolumn{2}{c}{$\Delta$L} & \multicolumn{1}{c}{L} & U & 
	\multicolumn{1}{c}{B} & \multicolumn{2}{c}{$\Delta$L} & \multicolumn{1}{c}{L} & U & UB  & \multicolumn{1}{c}{Tilt}\\
	  \hline
		DAO &  4 & +09 & 72 & W & 345 &  7.5 & +08.97 & 71.94 & W & 344.94 &  7.9 & 3.5 & 18.94\\
		AX  &  1 & +34 & 70 & W & 343 &   -  & +32.84 & 69.48 & W & 342.48 &   -  &  -  &   -  \\
		BXO &  2 & +30 & 63 & W & 336 &  4.0 & +29.75 & 62.84 & W & 335.84 &  4.9 & 2.2 & 25.10\\
		FKC & 43 & +21 & 32 & W & 305 & 15.0 & +20.85 & 32.86 & W & 305.86 & 14.7 & 9.3 & 24.25\\
		DKC & 29 & +27 & 03 & W & 276 & 10.0 & +26.69 & 02.48 & W & 275.48 &  9.7 & 6.5 & 26.35\\
		CRO &  6 & +17 & 09 & E & 264 &  4.5 & +16.88 & 09.54 & E & 263.46 &  4.7 & 1.5 & 10.92\\
		FKI & 43 & +26 & 11 & E & 262 & 18.0 & +25.44 & 11.40 & E & 261.96 & 18.6 & 9.0 & 18.30\\
		EHI &  9 & +18 & 55 & E & 218 & 14.0 & +17.27 & 54.47 & E & 218.53 & 13.3 & 3.2 &  1.45\\
		?   &  9 & +22 & 74 & E & 199 & 11.0 & +21.61 & 75.22 & E & 197.78 & 12.6 & 2.7 &  9.80\\
		CRO &  6 & -13 & 63 & W & 336 &  6.5 & -13.21 & 63.83 & W & 336.83 &  6.8 & 1.6 & 13.59\\
		FKC & 22 & -33 & 51 & W & 324 & 17.0 & -32.68 & 50.85 & W & 323.85 & 16.8 & 4.9 & 12.65\\
		DAC & 16 & -21 & 08 & E & 265 &  5.5 & -21.35 & 08.20 & E & 264.80 &  5.8 & 2.2 & 12.15\\
		BXO &  9 & -23 & 15 & E & 258 &  4.0 & -23.30 & 14.47 & E & 258.53 &  4.0 & 1.9 & 32.29\\
	  \hline  \hline
	\end{tabular*}
	\label{T-test}
  \end{center}
    \vspace{-0.2cm}
	$\it S$: Spot number, $\it B$: Latitude, $\Delta \it L$: Apparent Longitude, $\it L$: Longitude, $\it U$: Length of the group, $\it UB$: Latitudinal Length, $\it Tilt$: Inclination of the group with respect to the solar equator.\\
\end{table}

\begin{figure}[t]    
  \centerline{
	\includegraphics[width=0.6\textwidth,clip=]{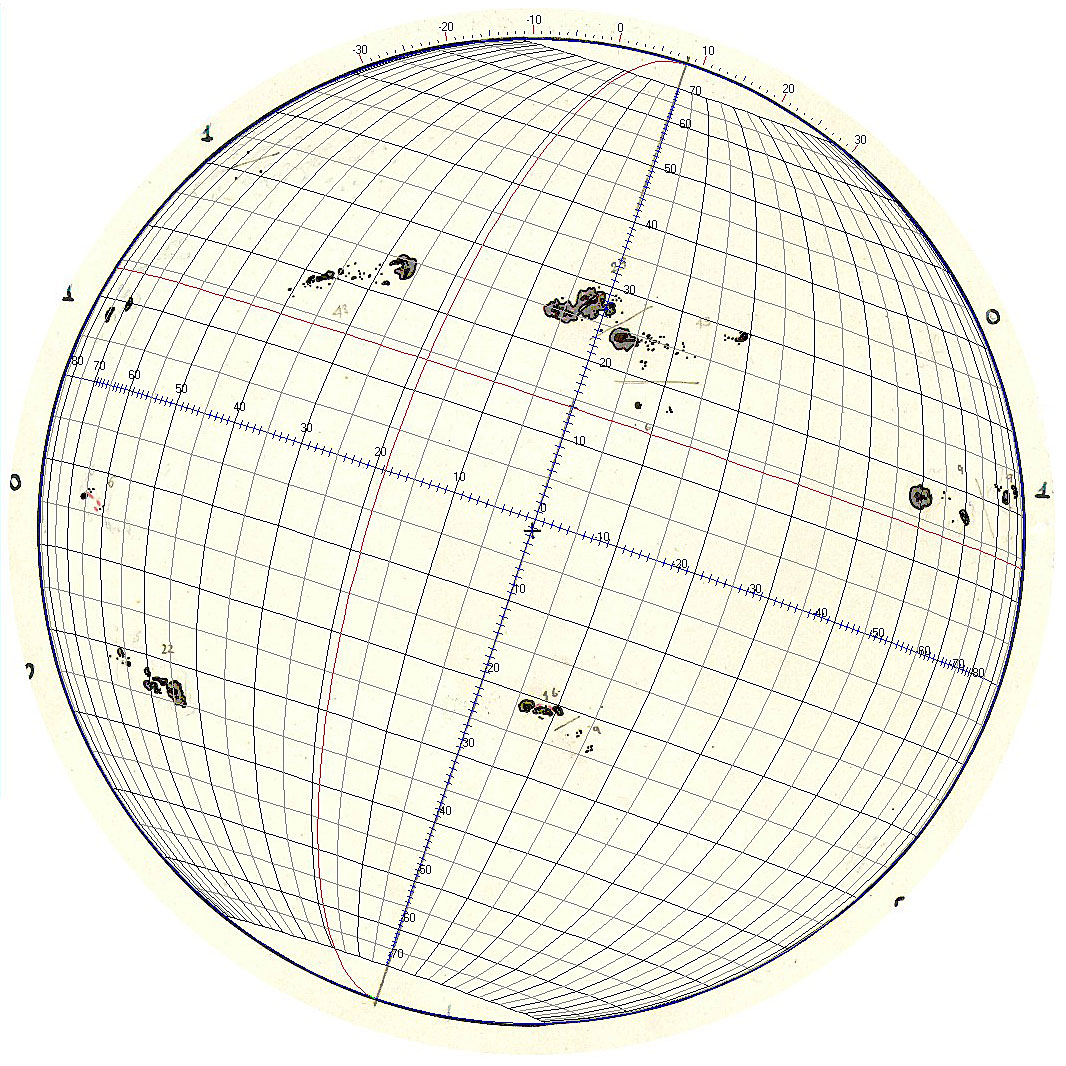}
  }
  \caption{Solar drawing of the observatory taken in 1989.12.19. Stonyhurst disk is superimposed for the time of the observation, 09:45 UT.}
  \label{F-figures-10}
\end{figure}

\section{Discussion} 
      \label{S-discussion}  
Nowadays, i.e. in the computer age, observation techniques, calculation steps and archiving the obtained knowledge are important and essential. From this perspective, the main purpose of CAMS is the shortening the time spent on processing the observation and to eliminate possible observer errors from the calculations. Second purpose is calculating the heliographic coordinates, latitudinal and longitudinal length, tilt angle and the rectangular area of the sunspot groups more precisely than manual measurements. Therefore, the measured values are saved with two decimals. When working on the positions of the sunspot groups, increasing the numerical precision in the coordinates gives a more accurate approach in the calculations. For example, in the butterfly diagram, the distributions of the sunspot groups with latitudes will appear to be more accurate instead of being concentrated on the latitude lines. 
\begin{figure}[t]    
  \centerline{
	\includegraphics[width=0.48\textwidth,clip=]{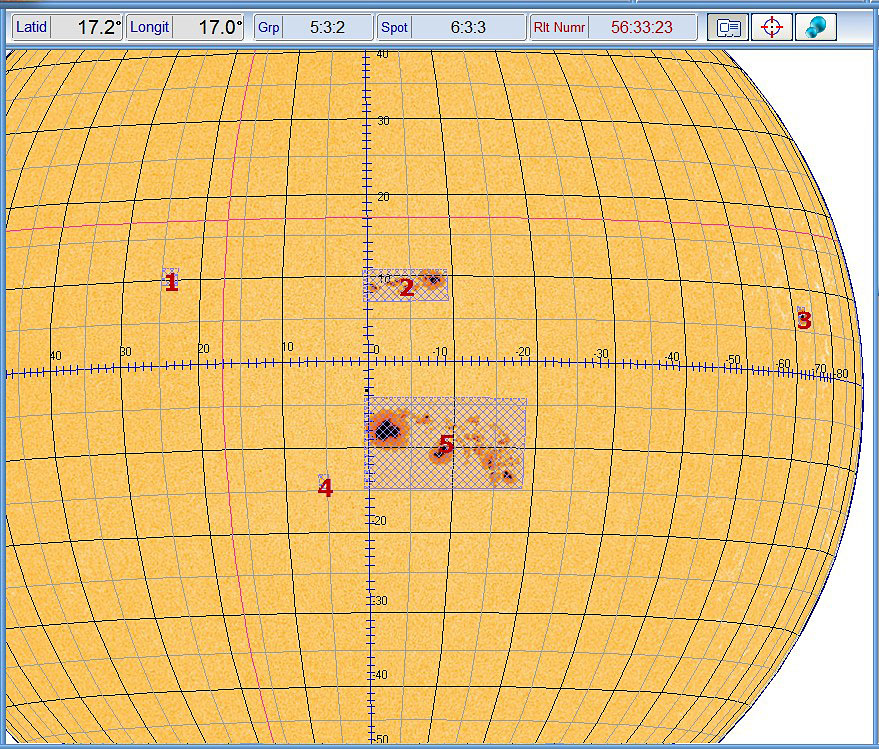}\hspace{0.2cm}
	\includegraphics[width=0.48\textwidth,clip=]{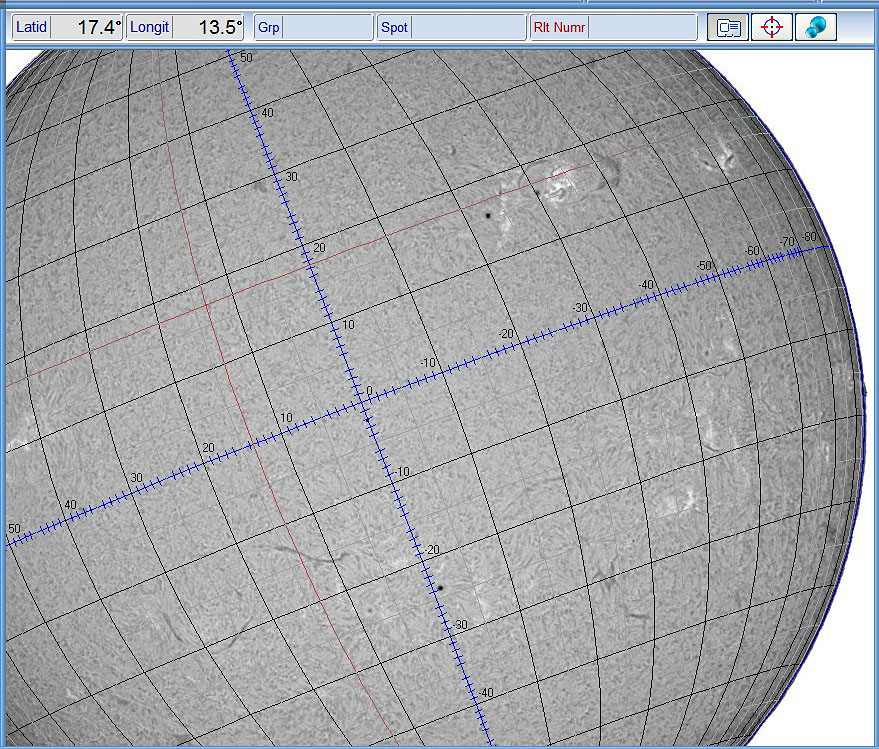}
  }
  \caption{Examples of the using digital solar image with CAMS. {\it Left}: A HMIIF (Helioseismic and Magnetic Imager Intensitygram - Flattened) image of the Solar Dynamics Observatory (SDO). {\it Right}: A chromospheric H-Alpha image taken with CCD at the Istanbul University Observatory.}
  \label{F-figures-11}
\end{figure}

Another point of the CAMS is that the Stonyhurst disks are drawn at the true angle of the latitude angle instead of the nearest integer value. That is to say, it is drawn at the angle, for example, 7.12$^\circ$ or 7.25$^\circ$ or 7.57$^\circ$ or 7.92$^\circ$, so not just for 7.0$^\circ$. The same is valid for the position angle, as well. Although CAMS is designed to process daily solar drawings, any digital solar image can be used by the program. Two examples of this are shown in Fig. 11. One is a HMIIF (Helioseismic and Magnetic Imager Intensitygram - Flattened) image of the Solar Dynamics Obsevatory, and the other is the chromospheric H-Alpha image taken with CCD at the university observatory. The heliographic coordinates of the sunspot groups can be measured more precisely with respect to the drawings. This can be done to obtain more accurate results or to eliminate the drawing errors of the observers in their studies.

The development of CAMS is still continuing and the integration with the database is under development. After this is done, it will be possible to obtain any group's evolution separately and all statistical information about the positions and types of the sunspot groups. Also the daily, monthly changes and general trend of the sunspot relative number can be easily generated graphically. Since the relative numbers are archived by making a distinction between the Northern and Southern, the variations of the relative number in both will be also examined separately. After sufficient data is gathered, the whole solar sunspot cycle will be analyzed in detail.

\begin{acknowledgements}
Thanks to Dr. Tuncay \"{O}z{\i}\c{s}{\i}k from the Turkish National Observatory to develop the CAMS and Assoc. Prof. Dr. Nurol Al Erdo\u{g}an from the Istanbul University  Science Faculty, Astronomy and Space Sciences Department for the idea to prepare and publish this article. Also, thanks to Asst. Ba\c{s}ar Co\c{s}kuno\u{g}lu for his contributions towards improving the language of the manuscript.  Also, thanks to anonymous reviewer for his/her valuable suggestions and comments improving manuscript This work supported by the Istanbul University Scientific Research Projects Commission with the project number 24242.
\end{acknowledgements}



\end{document}